# Performance and Degradation of A Lithium-Bromine Rechargeable Fuel Cell Using Highly Concentrated Catholytes

Peng Bai[1,*] and Martin Z. Bazant[1,2,*]

[1]Department of Chemical Engineering, Massachusetts Institute of Technology, Cambridge, MA 02139, USA

[2]Department of Mathematics, Massachusetts Institute of Technology, Cambridge, MA 02139, USA

* ISE Member

**Abstract:** Lithium-air batteries have been considered as ultimate solutions for the power source of long-range electrified transportation, but state-of-the-art prototypes still suffer from short cycle life, low efficiency and poor power output. Here, a lithium-bromine rechargeable fuel cell using highly concentrated bromine catholytes is demonstrated with comparable specific energy, improved power density, and higher efficiency. The cell is similar in structure to a hybrid-electrolyte Li-air battery, where a lithium metal anode in nonaqueous electrolyte is separated from aqueous bromine catholytes by a lithium-ion conducting ceramic plate. The cell with a flat graphite electrode can discharge at a peak power density around 9mW cm$^{-2}$ and in principle could provide a specific energy of 791.8 Wh kg$^{-1}$, superior to most existing cathode materials and catholytes. It can also run in regenerative mode to recover the lithium metal anode and free bromine with 80-90% voltage efficiency, without any catalysts. Degradation of the solid electrolyte and the evaporation of bromine during deep charging are challenges that should be addressed in improved designs to fully exploit the high specific energy of liquid bromine. The proposed system offers a potential power source for long-range electric vehicles, beyond current Li-ion batteries yet close to envisioned Li-air batteries.

# 1. Introduction

Li-ion batteries have powered the revolution in portable electronics and tools for decades, but their initial penetration into the market for electrified transportation has so far only achieved products that are very expensive and short in driving range [1]. Lithium-air batteries are considered among the most promising technologies beyond Li-ion batteries [2-4], since the very high theoretical specific energy may reduce the unit cost down to less than US$150 per kWh, while increase the driving range of an electric vehicle to more than 550km [5]. However, just as that Li-ion technology experienced many problems at its advent decades ago, Li-air technology is currently facing several challenges [6]. For nonaqueous Li-air batteries composed of lithium metal, organic electrolyte and porous air electrode, a robust electrolyte resistant to the attack by the reduced $O_2^-$ species is yet to be developed to enable highly reversible cycling [4, 5]. For aqueous and hybrid Li-air batteries that adopt solid-state electrolytes to protect the nonaqueous electrolyte and lithium metal anode from contamination, it is still quite challenging to improve the poor kinetics of the oxygen reduction reaction (ORR) and the oxygen evolution reaction (OER) simultaneously [7] and economically [4]. To circumvent this challenge, Goodenough et al [8-10] and Zhou et al [11] independently extended the hybrid Li-air battery to hybrid Li-redox flow batteries by flowing through liquid catholytes instead of air [12, 13]. The key concept of flowing electrodes is also exploited in semi-solid flow batteries [14], redox flow li-ion batteries [15, 16], and flowable supercapacitors [17].

One of the most attractive features of flow batteries is the decoupling of power and energy, which enables more flexible system customization, either by increasing the number of electrode pairs for higher power output, or by increasing the size of the tank and concentration of electrolytes to store more energy [13]. For electric vehicles with limited on-board space to store electrolytes, high solubility of the active species becomes especially important. Recognizing that iodine has an extremely high solubility in iodide solutions, Byon et al investigated the performance of dilute iodine/iodide catholyte in hybrid-electrolyte lithium batteries both in the static mode [18] and the flow-through mode [19], in which the end-of-discharge product is LiI. Concentrated iodine/iodide solution was also employed in a recent zinc-polyiodide flow battery, producing $ZnI_2$ at the end of discharge [20]. Comparing these two reports, although LiI and $ZnI_2$ solutions have similar capacity at their solubility limits, the use of a lithium anode increases the voltage almost three-fold, thus providing much higher specific energy. **Table 1** summaries the theoretical specific energies of catholytes used in several state-of-the-art flow or static-liquid batteries, where LiBr solution emerges as the best candidate, having almost twice the specific energy of the aqueous Li-air battery using alkaline catholyte (LiOH).

This extraordinary property has started to attract the attention of researchers to develop various Li-Br batteries. Such systems always involve a liquid-solid-liquid hybrid electrolyte, in order to accommodate the nonaqueous and aqueous electrolytes. During discharge, lithium metal in the nonaqueous electrolyte is oxidized into lithium ions ($Li \rightarrow Li^+ + e^-$), which migrate toward the cathode, while electrons travel through the external circuit to reach the cathode. At the surface of cathode, bromine is reduced by the incoming electrons to bromide ions ($Br_2 + 2e^- \rightarrow 2Br^-$), followed by fast complexation with bromine to form more stable tribromide ions ($Br^- + Br_2 \leftrightarrow Br_3^-$). The reactions are reversed during recharging. Zhao et al. fabricated a static Li-Br battery starting with 1M KBr and 0.3M LiBr solution, which was charged to 4.35V then discharged at various electrochemical conditions [23]. The maximum power it could deliver within the safety window was 1000W kg$^{-1}$, equivalent to 5.5mW cm$^{-2}$ if calculated with their loading density of LiBr (5.5mg cm$^{-2}$). Chang et al. paired a protected lithium metal anode [26] with a small glassy carbon electrode (3mm diameter) to test the performance of 0.1M Br$_2$ in 1M LiBr and 1M Br$_2$ in 7M LiBr solutions, respectively. The latter provided a peak power density of 29.67 mW cm$^{-2}$ at ~2.5V [27]. In the development of a better Li-Br battery, Takemoto and Yamada found that degradation of LATP plate is the major source of deterioration of the cell performance. Their careful analyses on samples soaked in dilute bromine/bromide solutions for 3 days suggested the development of a Li-ion depletion layer in LATP [28].

Given the strongly fuming and oxidative nature of bromine, it is understandable that previous work has only considered dilute electrolytes. Indeed, the high vapor pressure of bromine that builds up in a closed static liquid cell can easily rupture the LATP separator. Such problems can be avoided in a flow cell, but a practical way of utilizing the high specific energy of lithium-bromine chemistry has yet to be proposed and demonstrated, using highly concentrated bromine/bromide catholytes.

In this paper, we design and fabricate a lithium-bromine fuel cell; explore the feasibility of using highly concentrated bromine catholytes of six different compositions of LiBr and Br$_2$, representing different states of charge (SOC) associated with 11M LiBr solution by conservation of elemental bromine; and examine the degradation of the rate-limiting component, the lithium ion conducting solid electrolyte, by scanning electron microscopy and electrochemical impedance spectroscopy. Our results suggest that a properly designed rechargeable Li-Br fuel cell system has the potential to power long-range electric vehicles.

## 2. Experimental Section

*Fuel Cell Design and Fabrication*: The structure of the fuel cell is schematically shown in **Figure 1**, which is similar to the hybrid aqueous Li-air battery [29], where lithium metal in nonaqueous electrolyte is separated from aqueous catholytes by a solid electrolyte ($Li_2O$-$Al_2O_3$-$SiO_2$-$TiO_2$-$GeO_2$-$P_2O_5$, LATP, $10^{-4}$ S cm$^{-1}$, 25.4-mm square by 150-μm thick, Ohara Inc. Japan). A catalyst-free flat graphite plate is used as cathode. Catholytes flow through the cathode channel to complete the liquid-solid-liquid ionic pathway between lithium metal anode and graphite cathode. Details of the materials, design and fabrication of the fuel cell can be found elsewhere [30].

*Catholytes Preparation*: Theoretically, the fully discharged catholyte should not contain any $Br_2$ for further reduction reaction. It therefore must be pure LiBr solution. To avoid unexpected precipitation due to temperature fluctuations, we chose not to use the saturated LiBr solution (close to12M), but the slightly more dilute option, 11M LiBr aqueous solution, as the end-of-discharge catholyte. And according to the conservation of elemental bromine, we prepared 1M $Br_2$ in 9M LiBr (1M/9M), 2M/7M, 3M/5M, 4M/3M and 5M/1M solutions as the intermediate catholytes. Note that only 5M/1M solution has precipitated liquid $Br_2$ at the bottom of the solution, since the saturated concentration of $Br_2$ in 1M LiBr solution is around 2.2 M (1.93g $Br_2$ in 10ml LiBr solution). Only the supernatant solution, i.e. ~2.2M/1M, was used in the tests, but we nonetheless keep the notation of 5M/1M for easier understanding of its relation with other catholytes.

*Electrochemical Measurements*: Polarization curves were obtained by an Arbin battery tester (BT-2043, Arbin Instruments) at the flow rate of 1 ml min$^{-1}$ cm$^{-2}$. Every data point came from the averaged voltage of five-minute charge or discharge. Before testing a different catholyte, DI water and air were pumped to flush the tubing and cell at 5 ml min$^{-1}$ cm$^{-2}$ for 30 mins and 10 minutes, respectively. Potentiostatic EIS experiments of the Pt|LATP|Pt dry cells were conducted with Gamry Reference 3000, with a 5mV excitation from 0.1 Hz to 1MHz.

## 3. Results

## 3.1. Electrochemical Performance

The polarization curves shown in **Figure 2** reveal the linear relationship between the response voltages and the applied current densities. A peak power density of 8.5 mW cm$^{-2}$ at 1.8V can be obtained with 1M Br$_2$ in 9M LiBr (1M/9M) solution, which is consistent with the recent reports of both the static [23] and flow [30] Li-Br cells using *dilute* bromine catholytes. The fact that increasing the concentration of Br$_2$ here does not improve the discharge performance further confirms that the rate-limiting process is not transport in the liquid catholyte, but the conduction of lithium ions through the ceramic solid electrolyte. Data in Figure 2 also reveal that the slope of the polarization curves becomes increasingly steeper and power density smaller over time. This is due to the cumulative corrosion of the LATP electrolyte plate, consistent with the sequence of experiments from low Br$_2$ concentration to high Br$_2$ concentration.

**Figure 3** shows polarization data for the charging processes with the proposed bromine/bromide catholytes and the 11M LiBr solution without any Br$_2$ (0M/11M). Again, the slight increase of the slope reflects the cumulative deterioration of the LATP plate, consistent with the sequence of the experiments. At a given current density, the charging overpotential increases with the increase of bromine concentration. Note that for the 5M/1M solution, the saturated concentration of bromine in 1M LiBr is around 2.2M, similar to 2M/7M solution, which is expected to yield similar performance. However, since the latter has a much higher concentration of the supporting salt LiBr, it results in a much lower overpotential than for the 5M/1M solution.

Limited by the solid electrolyte, the maximum current density can be obtained is too low to complete a charge-discharge cycle before the breakdown of the solid electrolyte plate due to corrosion, or the exhaust of the electrolyte due to leakage, since even only 10 ml highly concentrated catholyte requires hundreds of days to be converted electrochemically. Here, to evaluate the efficiency, we choose another figure of merit widely used in the field of flow batteries [24], voltage efficiency, defined as the ratio of the discharging voltage and the charging voltage at a given current density.

The voltage efficiencies at ±0.5 mA cm$^{-2}$ shown in **Figure 4** are in the range of 80%-90%, which reflect relatively small voltage hysteresis (0.67V in average), better than typical Li-air batteries at lower currents. Due to the sluggish kinetics of ORR and OER, the voltage hysteresis of nonaqueous Li-air batteries using carbon electrode is typically larger than 1V even for currents as small as 0.105mA cm$^{-2}$ [31]. Cells with gold-modified electrodes and novel electrolytes containing redox mediators can exhibit 1V hysteresis at a slightly higher current density 0.313 mA cm$^{-2}$ [32]. The hysteresis only becomes comparable with TiC electrode and electrolyte of 0.5M LiPF$_6$ in tetraethyleneglycol dimethylether (TEGDME) [33]. Reaction kinetics in aqueous Li-air battery are even worse, due to higher activation energy to cleavage the O-O bond, but the hysteresis can reduce to 0.75V by increasing the operation temperature to 60ºC [29]. In general, Li-air batteries do not allow high power operation since the insulating discharge product would shut down the battery due to conformal coating to the air electrode [34].

In contrast, our Li-Br fuel cell does not have this problem due to the extraordinary solubility of its discharge product LiBr (~12 mole per liter of solution, or 18.89 mole per kg of water). Yet the open design allows operation outside the electrochemical stability window to achieve higher power output, since the generated gas can be brought out of the cell with the flowing stream, instead of building up inside the cell to rupture the LATP separator. While the fairly rapid degradation of LATP in concentrated bromine catholytes precludes the demonstration of reversible cycling with concentrated bromine catholytes, superior Coulombic efficiencies have been achieved in other aqueous lithium flow batteries using dilute I$_2$/LiI solution[19] and dilute K$_4$Fe(CN)$_6$ solution [9].

### 3.2. Degradation of the Solid Electrolyte

The deterioration of LATP has been intensely investigated for applications to aqueous Li-air batteries with various solutions, including water [35], acidic solutions [36, 37], and basic solutions [38]. In a recent work, Takemoto and Yamada [28] investigated the surface structure of the aged LATP samples by grazing incident X-ray diffraction (GIXD) and attenuated total reflection Fourier transform infrared spectroscopy (ATR-FT-IR). However, phase impurities and chemical changes that had been observed in strong acidic solutions [36, 37] were not found in their bromine-bromide catholytes containing 1M elemental Br [28], even though Br$_2$ disproportionates in water to form several species including acidic HBrO and HBrO$_3$. The degradation was attributed then to the only remaining conjecture of a Li$^+$-depletion layer developed into the surface of LATP plate. Here, we immersed small pieces of LATP samples in the proposed concentrated catholytes (containing 11M elemental Br) as well as the nonaqueous electrolyte for two weeks, and then characterized them with scanning electron microscopy (SEM) and electrochemical impedance spectroscopy (EIS).

SEM images of the aged LATP plates are shown in **Figure 5**. The glassy surface of the new LATP plate is difficult to focus in SEM, as the fine and shallow cavities cannot produce as strong contrast as the aged plates, in which both the size and depth of the cavities are clearly increased after immersion in different solutions. What was not discovered before is that the surface, although it still looks flat, develops roughness and asperities that can become loose. In fact, we observed that chunks of material were blown off in the flow of the catholytes, which indicates that there existed significant corrosion well below the deep cavities observed on the surface. We then focus on the middle part of their cross sections, typically in the region 70μm away from either surface, i.e. the least-corroded part of the solid electrolyte. It is clear to see that the cross sections of the new plate and the one immersed in 11M LiBr solution look dense and uniform with continuous and smooth connections among grains. However, nanopores between grains can be easily identified in the sample immersed in 1M/9M solution. With increased concentration of bromine, the cross sections of the samples look much more rough and porous. Individual grains with little contact to their surroundings reveal the severe corrosion of the grain boundaries. The SEM images of the sample immersed in nonaqueous electrolyte also show deep cavities on the surface and rough and porous morphology in the bulk, consistent with earlier reports [30]. These structural degradations are well associated with the deterioration of the conductivity of the solid electrolyte, which can be evaluated quantitatively by electrochemical impedance spectroscopy.

To obtain the EIS spectra for all eight samples, shown in **Figure 6**, two pieces of platinum foil were attached to the anvils of a micrometer, which was used to hold the sample and form a Pt|LATP|Pt dry cell, see in Figure 6j. This simple design avoids short circuiting at edges of the small LATP samples created by sputtering gold electrode onto both surfaces. While it may not guarantee accurate measurements of the absolute conductivity of the LATP samples, due to less intimate contact than sputtered gold electrode, it is adequate for us to investigate the relative increase of the impedance of the aged LATP samples and compare them with the new LATP sample. Consistent with the SEM observation, the new LATP plate and the one soaked in 11M LiBr solution exhibit similar impedance behavior, but the latter forms a much clearer and smaller semicircle at high frequencies, indicating improved conductivity. In general, the impedance of the aged LATP plates increases with the increase of bromine concentration in the solutions. Note that for 5M/1M solution, the saturated concentration of bromine is around 2.2M, and its impedance spectra coincide with that of 2M/7M solution.

Various equivalent circuit models have been proposed to fit the impedance of ceramic solid electrolytes [35, 39, 40]. As shown in Figure 6i, we attribute the impedance to two parts, one from the grains and the other from the grain boundaries [40]. **Figure 7** shows the fitted resistances of grains and grain boundaries corresponding to the results displayed in Figure 6. Both the grain and grain-boundary resistance of the sample soaked in 11M LiBr are lower than the new plate, which coincide with the smooth cross section shown in Figure 5b. The resistances of other samples have a clear trend with respect to the concentration of dissolved bromine. The one soaked in nonaqueous battery electrolyte shows increased resistance similar to that soaked in 1M/9M solution, although the cross-section morphologies look quite different.

## 4. Discussion

The strong corrosion effects of bromine solution jeopardize the durability of the fuel cell. This difficulty necessitates a system design shown in **Figure 8**a. The fuel cell system could involve a primary fuel tank storing pure bromine, which can be released through an electronic valve into a secondary tank to maintain the optimal concentration of the catholyte that will be circulated through the fuel cell until the full tank of bromine is exhausted and completely converted to LiBr solution. For systems using currently available water-stable solid electrolytes, one may consider only using dilute bromine (but not necessarily dilute LiBr) catholytes, which could provide similar peak power [30] and better Coulombic efficiency and longer life [9, 19] as shown in previous work. Apparently, the combination of the flow cell and the $Br_2$ tank is the only way to exploit the high specific energy of lithium-bromine chemistry, since the lack of a strong and corrosion-resistant solid electrolyte implies that static Li-Br batteries will only work with limited amount of dilute bromine catholytes [23, 28], whose specific energy (~100Wh (kg-catholyte)$^{-1}$) is not superior to existing Li-ion batteries (~500Wh (kg-cathode)$^{-1}$), and cycle life not longer than Li-redox flow batteries using less corrosive catholytes [9, 11, 19].

While discharging with $Br_2$ catholyte is straightforward, the key to achieve the proposed theoretical specific energy and a high Coulombic efficiency relies on whether all the lithium ions and bromide ions generated during discharge can be recovered to metallic lithium and free bromine, respectively. We performed constant-voltage charging with saturated $Br_2$ in 1M LiBr solution, i.e. the supernatant solution in the 5M/1M catholyte, for 46 hours. The total charged capacity was 1.9 mAh, which should convert to 1 mL of liquid $Br_2$. However, the color of the catholyte at the end of the 46-hour charging is much lighter and does not fume as much as the initial catholyte, which indicates the loss of bromine by evaporation. Installing a $Br_2$ extractor, which can be as simple as an air blower plus a condenser [41], to separate the free bromine from the recharging stream may help reduce the energy loss by evaporation, and also alleviate the corrosion of LATP plate by keeping a low bromine concentration.

As demonstrated in section 3, the highly concentrated 11M LiBr solution is both the most efficient catholyte for charging and the least corrosive catholyte to the LATP plate. Therefore, using 11M LiBr solution as a standard charging catholyte and modularizing the 11M-LiBr tank with the bromine extractor off-board, while only keeping the discharging module on-board, may become a highly efficient mode of operation for electric vehicles. The off-board charging system could also be enlarged as a recharging/refueling station, where the recharging stream can be guided to and processed with more sophisticated extractors, and the extracted bromine refueled into the on-board tank. The situation is analogous to capturing the exhaust of a combustion engine and exchanging it for a fresh tank of gasoline at the station – with the important difference that exhaust product (11M LiBr solution to be returned) is efficiently converted back into chemical fuels (liquid $Br_2$ and Li metal to be picked up) at the station, using only electricity without directly consuming any chemicals. Since the electricity could come from a renewable resource (solar or wind) at the refueling station, this concept could provide a means of sustainable power for electrified transportation.

Just as with all other lithium metal batteries, dendritic electrodeposition of lithium during recharging is a serious safety concern and lifetime challenge. Using solid electrolytes is believed to be an effective method to block lithium dendrite from shorting the cell, but the water-stable LATP is unstable in contact with lithium metal, which is a reason for the nonaqueous buffer layer employed in our design. Developing composite solid electrolytes that provide dual stability against lithium metal and water could help solve this problem [42]. Directly stabilizing the lithium metal anode during high-rate cycling would also be helpful. Although not yet investigated in deep-recharging situations, recent advances suggested many promising technologies, including creating protection layer of carbon semispheres to isolate lithium deposition [43], using extremely highly concentrated organic electrolyte to retard the concentration instability at metal surfaces [44], adding halogen ions [45] or metal ions[46] to modulate the reactions, and modifying the surface charge of the separator to trigger stable "shock electrodeposition" [47, 48].

By exploiting the fast kinetics of aqueous bromine/bromide catholytes, the Li-Br fuel cell exhibits much better power density than state-of-the-art Li-air batteries, which usually discharge well below 3mW cm$^{-2}$ even with catalyzed electrodes and modified electrolytes [29, 31-34]. To achieve power densities comparable to proton exchange membrane (PEM) fuel cells already installed in electric vehicles, however, a thinner solid electrolyte with higher ionic conductivity, presumably supported by strong substrates, may need to be developed. Another approach could be to remove the rate-limiting solid electrolyte to fabricate a membraneless system [49,50], whose power density could be increased by orders of magnitude, as the ionic conductivities of the liquid electrolytes are at least two orders of magnitude higher than that of typical solid electrolytes [27].

## 5. Conclusion

We have designed and fabricated a rechargeable lithium-bromine fuel cell and investigated the feasibility of using highly concentrated bromine catholytes in order to exploit the very high specific energy of lithium-bromine chemistry. Our results reveal that the commercially available water-stable solid electrolyte LATP degrades quickly in the concentrated bromine catholytes, making long-time operation and cycling almost prohibitive. However, a new system design, which combines the fuel cell with a primary tank of pure liquid bromine and a secondary tank for dilute bromine/bromide catholytes may provide the possibility to eventually achieve the theoretical high energy density. While static Li-Br batteries are only able to work with limited amount of dilute catholytes, yielding less appealing performance in specific energy and cycle life than existing technologies, the proposed Li-Br system could be a viable technology to provide sustainable power for long-range electric vehicles, as research continues toward higher-power and more robust Li-air batteries.

**Acknowledgements**
This work was supported by a seed grant from the MIT Energy Initiative.

**Table 1.** Comparison of the specific energies of various fully discharged catholytes at their solubility limits.

| Discharge product | Solubility[21, 22] [g per 100ml of water] | Molality [mol kg$^{-1}$ of water] | Specific capacity [Ah kg$^{-1}$ of solution] | OCV [V] | Specific energy [Wh kg$^{-1}$ of solution] |
|---|---|---|---|---|---|
| LiBr | 164.00 | 18.89 | 191.72 | 4.13 [23] | 791.82 |
| LiI | 165.00 | 12.33 | 124.68 | 3.57 [18] | 445.12 |
| LiOH | 12.40 | 5.18 | 123.45 | 3.4 [2] | 419.73 |
| $ZnBr_2$ | 447.00 | 19.85 | 194.51 | 1.85 [24] | 359.84 |
| $ZnI_2$ | 332.00 | 10.40 | 129.06 | 1.30 [20] | 167.77 |
| $FeCl_2$ | 62.50 | 4.93 | 81.33 | 4.06 [11] | 330.19 |
| $K_4Fe(CN)_6 \cdot 3H_2O$ | 28.00 | 0.66 | 13.88 | 3.99 [10] | 55.38 |
| $Li_2S_n$ | --- | --- | --- | --- | 170[25] |

Figure 1

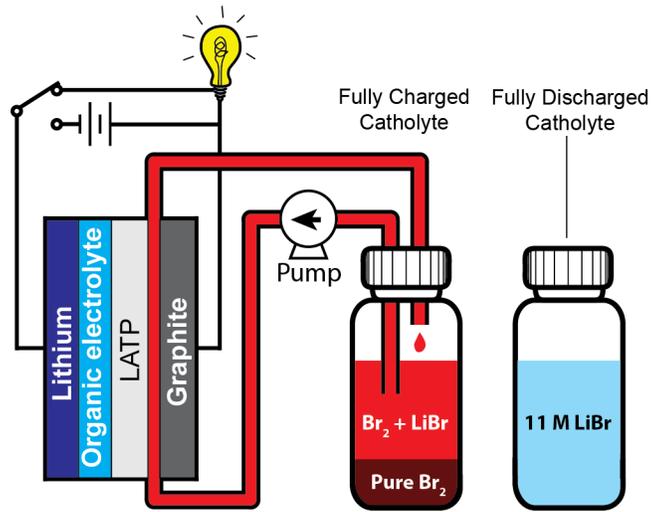

**Figure 1.** Schematic illustration of the Li-Br fuel cell.

Figure 2

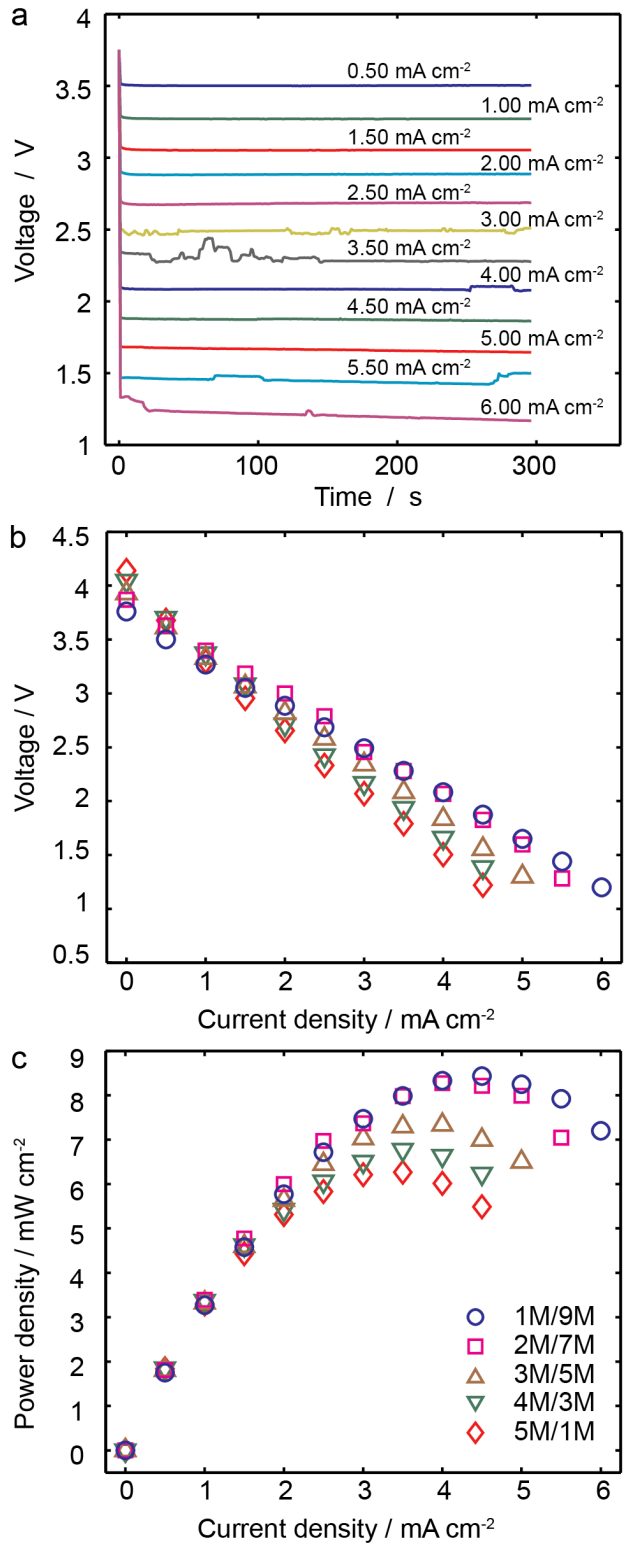

**Figure 2.** (a) 5-min Galvanostatic discharging with 1M/9M catholytes. (b) Polarization curves of the averaged voltages versus the applied current densities and (c) the corresponding power output for the proposed catholytes. Note that the saturated concentration of $Br_2$ in 1M LiBr is around 2.2M, which is the actual catholyte passing through the cell, no liquid $Br_2$ was directly introduced into the cell.

Figure 3

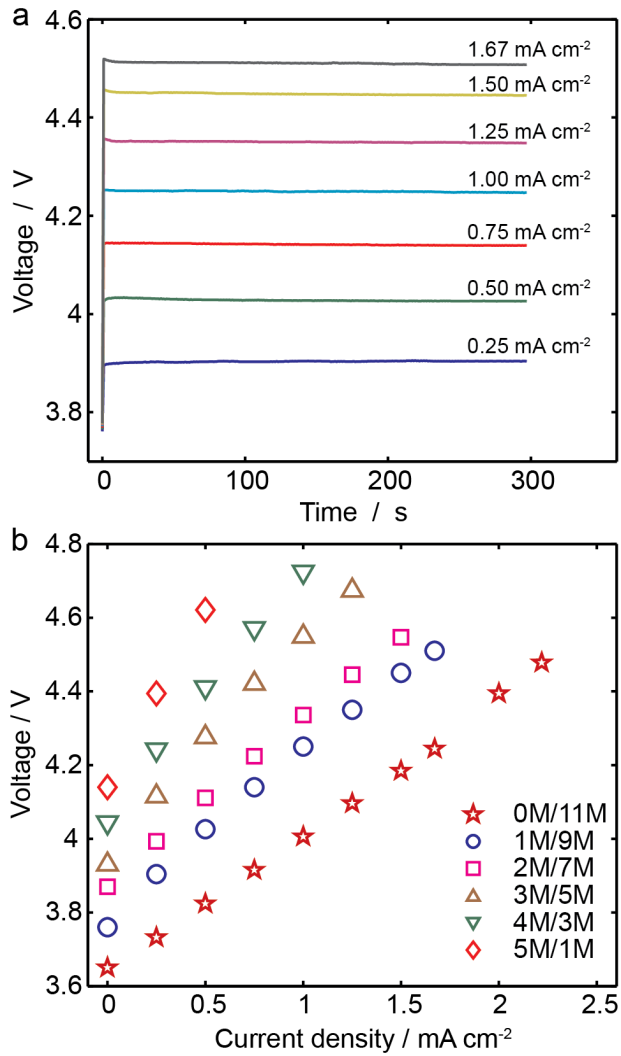

**Figure 3.** (a) 5-min Galvanostatic charging with 1M/9M catholytes. (b) Polarization curves of the averaged voltages versus the applied current densities for the proposed catholytes.

Figure 4

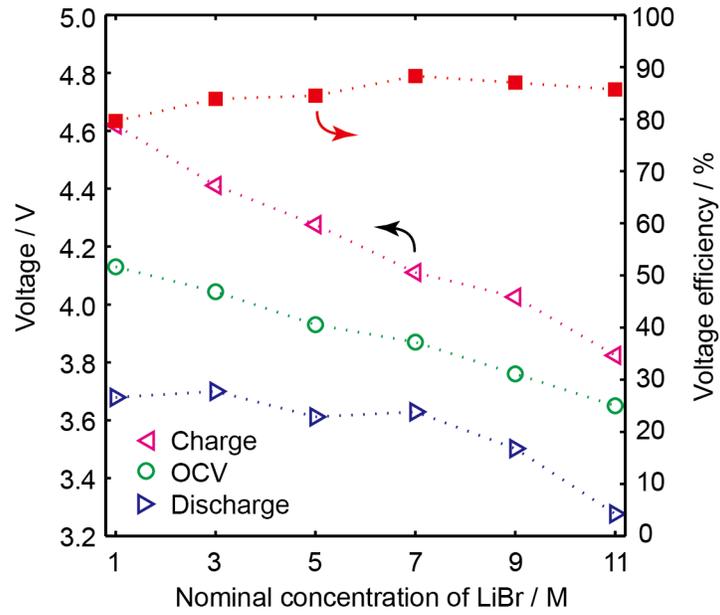

**Figure 4.** Open-circuit and the polarization voltages under ±0.5 mA cm$^{-2}$ with the corresponding voltage efficiencies for the series of catholytes.

Figure 5

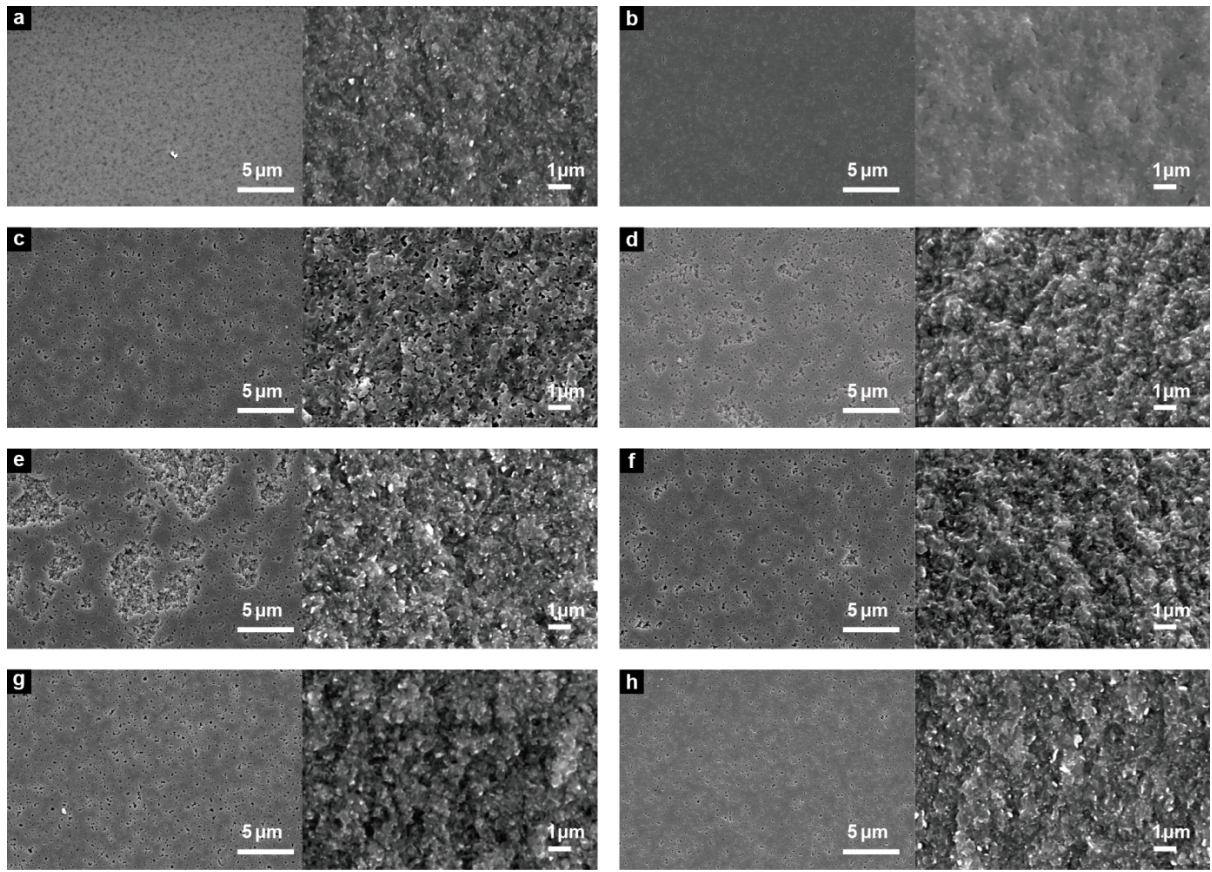

**Figure 5.** Scanning electron microscopy images revealing the morphologies of the surfaces (left) and cross sections (right) of the (a) new LATP plate, and those immersed in (b) 0M/11M, (c) 1M/9M, (d) 2M/7M, (e) 3M/5M, (f) 4M/3M, (g) 5M/1M catholytes and (h) nonaqueous electrolyte for two weeks.

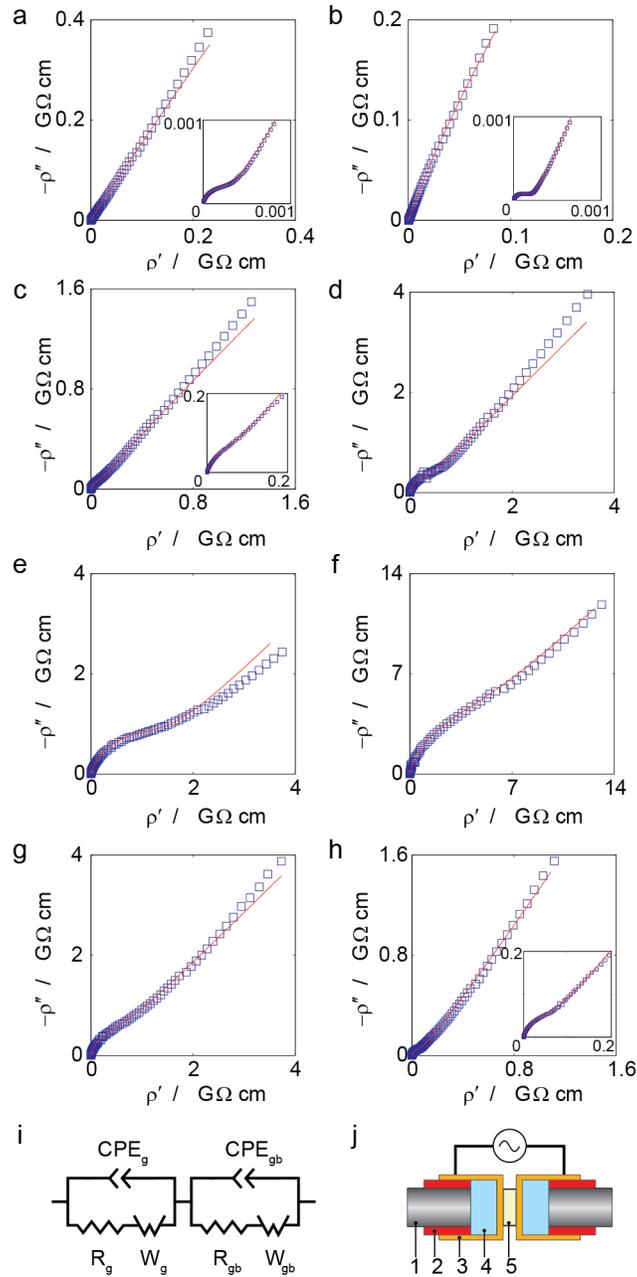

**Figure 6.** Electrochemical impedance spectroscopy spectra of the (a) new LATP plate, and those immersed in (b) 0M/11M, (c) 1M/9M, (d) 2M/7M, (e) 3M/5M, (f) 4M/3M, (g) 5M/1M catholytes and (h) nonaqueous electrolyte for two weeks. (i) Equivalent circuit model used to fit the experimental results. (j) Experimental setup, 1 – anvil of the micrometer, 2 – insulating layer, 3 – platinum electrode, 4 – glass substrate, 5 – LATP sample. Open squares are experimental

data, and the solid lines are fitting results. $\rho = Z \times A/l$, where Z is the measured impedance, A the surface area of the sample and l the thickness of the sample.

Figure 7

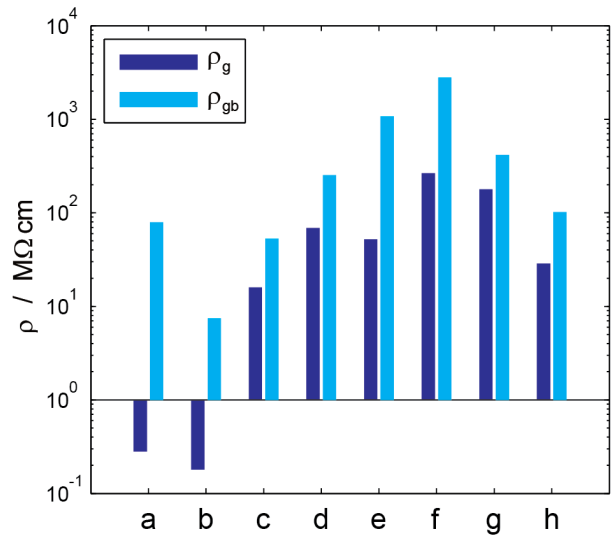

**Figure 7.** Effective resistivity of grains, $\rho_g = R_g \times A/l$, and grain boundaries, $\rho_{gb} = R_{gb} \times A/l$, from the impedance fitting in Figure 6.

Figure 8

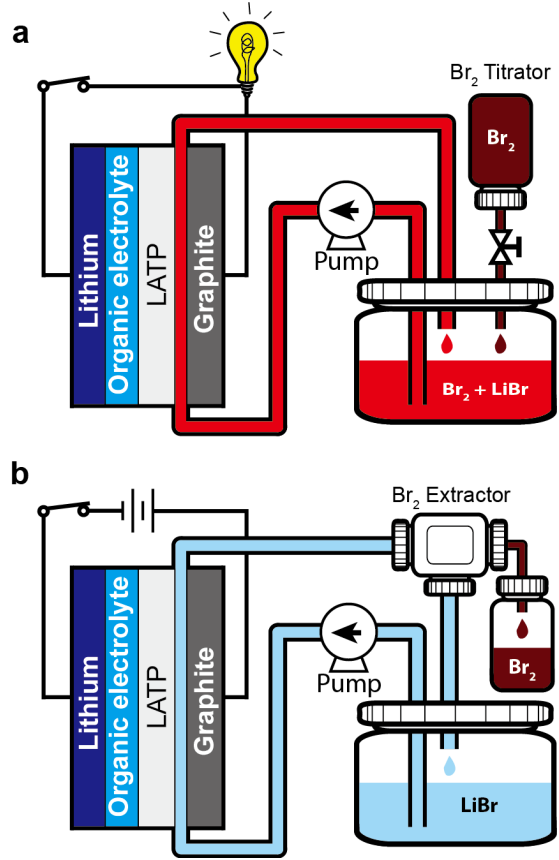

**Figure 8.** Schematic illustration of the rechargeable Li-Br fuel cell system. (a) Discharging mode with a Br$_2$ titration system to maintain the optimal concentration of bromine in the catholyte. (b) Regenerative mode with a Br$_2$ extractor to ensure a high efficiency by keeping a low bromine concentration in the catholyte.